\input harvmac
\skip0=\baselineskip
\divide\skip0 by 2

\def\tmpsp{\the\skip0}

\def\skipthis#1{{}}

\def\IR{\relax{\rm I\kern-.18em R}}
\def\IZ{\relax\ifmmode\hbox{Z\kern-.4em Z}\else{Z\kern-.4em Z}\fi}
\def\IQ{\relax{\rm I\kern-.40em Q}}
\def\IS{\relax{\rm I\kern-.18em S}}

\Title{\vbox{\baselineskip12pt\hbox{hep-th/0008142}\hbox{}
\hbox{HUTP-00/A035}}}{Superstrings and Topological Strings at Large N}

\centerline{Cumrun Vafa}
\bigskip\centerline{Jefferson Physical Laboratory}
\centerline{Harvard University}
\centerline{Cambridge, MA 02138, USA}

\vskip .3in \centerline{\bf Abstract}
We embed the large N Chern-Simons/topological string duality 
in ordinary superstrings.  This corresponds to a large $N$
duality between
generalized gauge systems with $N=1$ supersymmetry in $4$ dimensions and
superstrings propagating on non-compact Calabi-Yau
manifolds with certain fluxes turned on.  We also show that in a
particular
limit of the $N=1$ gauge
theory system,  certain superpotential terms in the $N=1$ system
(including deformations if spacetime is non-commutative)
 are captured to all orders in $1/N$ by
the amplitudes of non-critical bosonic strings propagating
on a circle with self-dual radius.
We also consider
D-brane/anti-D-brane system wrapped over vanishing
cycles of compact
Calabi-Yau manifolds and argue that at large $N$ they induce
a shift in the background
to a topologically distinct Calabi-Yau, 
which we identify
as the ground state system of the Brane/anti-Brane system.

\smallskip
\Date{August 2000}
\newsec{Introduction}

The idea that large $N$ gauge theories should have a phase
described by perturbative strings, set forth by `t Hooft
\ref\tho{G. `t Hooft, ``A Planar Diagram Theory for Strong
Interactions,'' Nucl. Phys. {\bf 72} (1974) 461.},
 has been beautifully realized by various examples.
The first example of this kind was found by Kontsevich \ref\konts{M.
Kontsevich, ``Intersection Theory on the Moduli Space of Curves and the
Matrix Airy Function,'' Comm. Math. Phys. {\bf 147} (1992) 1.}, which
relates the bosonic string theory coupled to certain matter
($(1,2)$ minimal model,
which is equivalent to pure topological gravity formulated
by Witten \ref\wittopg{E. Witten, ``On the Structure of the Topological
Phase of Two-Dimensional Gravity,'' Nucl. Phys. {\bf B340} (1990)
281.}), to a matrix integral
with cubic interaction (which can be viewed as
a particular gauge theory in zero dimensions)\foot{
This is not the same as the old matrix model which discretizes
the worldsheet--rather it is the target space description
exactly in line with `t Hooft's conjecture.}.
  Many more examples were also found in the context of non-critical
  bosonic strings.
For example, it was found \ref\disv{J. Distler and C. Vafa,
``A Critical Matrix Model at $c=1$,'' Mod. Phys. Lett. {\bf A6} (1991)
259.}\
that bosonic strings propagating
on a circle with self-dual radius is equivalent to Penner matrix model
\ref\penn{R.C. Penner, ``Perturbative Series and the Moduli
Space of Riemann Surfaces,'' UCSD preprint (1986).}.  

More recently it was recognized that `t Hooft's conjecture is also realized
even for much more complicated and
physically more interesting gauge theories \ref\malda{J. 
Maldacena, ``The Large N limit of Superconformal Field Theories
and Supergravity,'' Adv. Theor. Math. Phys. {\bf 2} (1998) 231.}\ref\gkp{
S.S. Gubser, I.R. Klebanov and A.M. Polyakov, ``Gauge Theory Correlators
From Non-critical String Theory,'' Phys. Lett. {\bf B428}
 (1998) 105.}\ref\wittads{E. Witten, ``Anti-de Sitter Space and Holography,''
 Adv. Theor. Math. Phys. {\bf 2} (1998) 253.}.
In particular certain gauge theories at large $N$
are equivalent to superstrings
propagating on AdS backgrounds.  Another example of a string/large $N$
duality was discovered in \ref\gopv{R. Gopakumar and C. Vafa,
``On the Gauge Theory/Geometry Correspondence,'' hep-th/9811131.},
 where it was
shown that large $N$ limit of Chern-Simons gauge theory
on $S^3$ is equivalent to topological strings on a non-compact
Calabi-Yau threefold 
which is a blow up of the conifold
(given by $O(-1)+O(-1)$ bundle over ${\bf P}^1$).
This duality was tested
to all orders in the $1/N$ expansion
including checks at the level of Wilson Loop
observables of the Chern-Simons theory \ref\oov{H. Ooguri
and C. Vafa, ``Kont Invariants and Topological Strings,''
Nucl. Phys. {\bf B577} (2000) 419.}\ref\mala{J.M.F. Labastida and M.
Marino,``Polynomial Invariants for Torus Knots and Topological Strings,''
hep-th/0004196.}.  It is also known \ref\peri{V. Periwal,
``Topological Closed-String Interpretation of Chern-Simons Theory,''
Phys. Rev. Lett. {\bf 71} (1993) 1295.}\ that in some limit
(large $N$, fixed Chern-Simons coupling $k$) this theory has the same
partition function
 as bosonic strings at the self-dual radius.

This paper was motivated by trying to connect the duality
discovered in \gopv\ with the dualities discovered
in the context of AdS/CFT correspondences.  The basic
idea is to consider type IIA superstring propagating in the conifold 
background (which
is symplectically the same as $T^*S^3$) in the presence
of $N$ D6 branes wrapped around $S^3$ and filling the
spacetime.  It has been known
\ref\bcovii{M. Bershadsky, S. Cecotti, H. Ooguri
and C. Vafa, ``Kodaira-Spencer Theory of Gravity
and Exact Results for Quantum String Amplitudes,''
Comm. Math. Phys. {\bf 165} (1994) 311.}\ 
that the topological string amplitudes
for the internal theory on the non-compact Calabi-Yau
compute superpotential terms on the left-over $R^4$ worldvolume of 
the $D6$ brane.   On the other hand it is also known that the
internal topological string theory with $N$ D-branes wrapped on $S^3$
is equivalent to Chern-Simons gauge theory on $S^3$ \ref\wittcstop{E. Witten,
``Chern-Simons Gauge Theory as a String Theory,'' hep-th/9207094.}. 
  Thus the duality found in \gopv\ suggests
that type IIA string on the conifold with $N$ D6 branes is equivalent
to the blown up version of the conifold with no branes left over.
At first sight this sounds strange, because having no D-branes
left would naively suggest a theory with $N=2$ supersymmetry
rather than $N=1$.  Moreover Ramond
fluxes should also be turned on
 in the blown up geometry corresponding to the flux
generated by the
D6 brane.  The main puzzle was why in the dual
topological string theory discovered in \gopv\ there is
no mention of RR fluxes?  Indeed it is an ordinary topological
string (the A-model) on the blown up conifold.

The resolution turns out to be that turning on the RR flux does not affect
the topological string amplitudes, and the dual string theory
{\it does} involve RR fluxes.  
Turning on RR flux, however, does generate an $N=1$ superpotential
term \ref\tayv{T.R. Taylor and C. Vafa,``RR Flux on Calabi-Yau and
Partial Supersymmetry Breaking,'' Phys. Lett. {\bf B474} (2000)
 130.}\ref\mayr{P. Mayr, ``On Supersymmetry Breaking in String
 Theory and its Realization in Brane World,'' hep-th/0003198.}, which can be computed
in terms of the topological string amplitudes.
Thus the duality found in \gopv\ can be
viewed as an all order check in the $1/N$ expansion for 
the $N=1$ superpotential
computations in the context of this type IIA
superstring/gauge theory duality.  One can also consider
the mirror symmetry acting on all these statements, which
as noted in \gopv\ give rise to similar dualities.  In 
the superstring realization, the mirror case (in a certain limit)
 would correspond to considering type IIB string on the blow
up of the conifold with $N$ D5 branes wrapped on ${\bf P}^1$
and we end up with type IIB on deformed conifold geometry $T^*S^3$
but with RR flux turned on.

One can also consider wrapped D-brane in the context
of compact Calabi-Yau manifolds.  However in this case we
also need to put anti-D-branes, in order to have no net
D-branes.  In this case we conjecture that the large $N$ limit
will correspond to having a new Calabi-Yau with fluxes, which can decay
as discussed in \ref\bupo{R. Bousso and J. Polchinski,
``Quantization of Four-Form Fluxes and Dynamical
Neutralization of the Cosmological Constant,'' hep-th/0004134.}\
to a theory with no fluxes left-over and with
supersymmetry increased 
to $N=2$.  The effect of the non-BPS states has
been
to shift the background to a new background.  This is a novel
way of deforming backgrounds, and as we will suggest
later in the paper may have many interesting extensions.

The organization of this paper is as follows:
In section 2 we review aspects of topological string amplitudes
and what they compute in the corresponding superstring theory.
In section 3 we revisit the duality of \gopv\ and embed
it in the context of Type IIA superstrings.  In section 4
we apply mirror symmetry to the statements in section 3
and discuss the equivalent Type IIB superstring theory.
In section 5 we discuss possible applications of $c=1$
non-critical bosonic strings to the question of generation
of superpotential  in the large
$N$ limit of $N=1$ supersymmetric gauge theory.
In section 6 we discuss wrapped brane/anti-brane systems in the context
of compact Calabi-Yau manifolds and use the
above duality to make new predictions about the shift in the background.
In section 7 we discuss some generalizations of this work.

While preparing this paper, three papers appeared which
have overlaps with different aspects of our work.  In particular
\ref\klst{I. Klebanov and M.J. Strassler,``Supergravity and a Confining
Gauge Theory: Duality Cascades and $\chi$SB-Resolution of Naked
Singularities,'' hep-th/0007191.}\ref\maldn{J. Maldacena and
C. Nunez, ``Towards the Large N Limit of Pure $N=1$ Super Yang-Mills,''
hep-th/0008001.}\
have some overlap with our work in the context of 
large $N$ duals of $N=1$ gauge theories in the context of
type IIB strings,
which we will briefly comment
on in section 4. 
Also the same configuration of wrapped D-branes/anti-D-branes
considered in section 6 was also studied in
\ref\mbd{T. Banks, M. Dine and L. Motl,
``On Anthropic Solutions of the Cosmological Constant Problem,''
hep-th/0007206.}\  in a different context.

\newsec{Topological Strings and Superstrings}

In this section we discuss aspects of topological strings
and their relevance for superpotential computations in the
corresponding superstring compactifications.  We will divide
our discussion to two parts: Closed string case (i.e. without
D-branes) and open string case (i.e. including D-branes). 
We also point out the relevance of topological
string amplitudes for $N=1$ superpotential
computations when RR-fluxes are turned on.

\subsec{Closed topological string and superstring amplitudes in 4d}
 Consider A-model topological strings on a Calabi-Yau manifold $K$
(similar remarks apply to the mirror B-model).  For simplicity
of notation
let us assume that the CY manifold has only one Kahler class,
parameterized
by the complexified Kahler parameter $t$. 
Then closed topological
string amplitude on $K$ is given by
\eqn\tops{F(t,{\lambda_s})=\sum_g\lambda_s^{2g-2}F_g}
$$F_g= \sum_{d} F_{d,g} e^{-dt}$$
where, roughly speaking $F_{d,g}$ denote the ``numbers''
(Gromov-Witten invariants)
of genus $g$ curves in class $d$.
The topological
strings compute certain amplitudes in the corresponding 
type IIA superstring compactifications on the Calabi-Yau
\bcovii\ref\naret{I. Antoniadis, 
E. Gava, K.S. Narain and T.R. Taylor,
``Topological Amplitudes in String Theory,''
Nucl. Phys. {\bf B413} (1994) 162.}\ref\berkv{N. Berkovits
and C. Vafa, ``N=4 Topological Strings,''
Nucl. Phys. {\bf B 433} (1995) 123.}. In particular
they compute terms in the action of the form
$$\int d^4\theta {\cal W}^{2g} F_g(t)=gR^2 F^{2g-2}F_g(t)+...$$
where ${\cal W}_{\alpha \beta}$ denotes the graviphoton field strength
multiplet, $R^2$ and
$F^{2g-2}$ denote certain contractions of the self-dual part
of the Riemann tensor and of the gravi-photon field strength, 
and $t$ denotes the vector superfield
with the vev of the lowest component being the Kahler parameter $t$.
One way to derive this formula is to note that with $2g-2$
insertions of the spin operator, needed to compute
the amplitude involving the $F^{2g-2}$,
the ordinary sigma model is topologically twisted.  At genus $0$
what one gets is 
$$\int d^4 \theta F_0(t)=\partial ^2F_0(t) F^t\wedge F^t+...$$
where $F^t$ denotes the (self-dual part of the) $U(1)$
field strength in the same multiplet as $t$.  In the type IIA
this arises from the 4-form field strength $G$ by setting it to
$$G=F^t\wedge \omega_t$$
where $\omega _t$ denotes the Kahler form associated to $t$.

It is natural to ask what changes in the closed
topological string computations when we turn on some RR
flux in the target space.  The choices are\foot{
We can also include the 0-form field strength
dual to 10 form field strength in type IIA, but
since we will not deal with it in this paper we will not
discuss it.  It will give rise to an $N=1$ superpotential
of the form $\int G_0 \wedge k^3$.} the 2-form
field strength in the internal space $F$, 4-form field
strength $G_{int}$ along the internal CY directions and the $G$ along
the spacetime directions $G_4$, which we
equivalently study in terms of the dual 6-form field
strength $G_6=*G_4$.  It turns out that
the
topological
string amplitudes in the presence of RR fields is
not modified at all!
This is particularly simple to show
in the Berkovits formalism \ref\berko{N. Berkovits,
``Covariant Quantization of the Green-Schwarz
Superstring in a Calabi-Yau Background'', Nucl. Phys. {\bf B431} 
(1994) 258.}\berkv \ref\berksi{
N. Berkovits and W. Siegel,
``Superspace Effective Actions for 4D Compactifications
of Heterotic and Type II Superstrings'', 
Nucl. Phys. {\bf B462} (1996) 213.}.
Instead of demonstrating it in this way we follow
a related idea, which we will need later in this paper, by 
studying the
generation of $N=1$ superpotential
terms in the presence of RR fluxes, which we will discuss next.

\subsec{Generation of superpotential due to internal field strength}

RR fluxes have been studied in the context of
CY compactifications \ref\pols{J. Polchinski and A. Strominger,
``New Vacua for Type II String Theory,'' Phys. Lett. {\bf B388}
(1996) 736.}\ref\Mich{J. Michelson, ``Compactifications
of Type IIB Strings to Four Dimensions with
Non-trivial Classical Potential,'' Nucl. Phys.
{\bf B495} (1997) 127.}\tayv\mayr. 
  In particular it
 has been shown in \tayv\mayr\ that turning
on internal field strength in the CY leads to generation
of superpotential terms in 4d $N=1$ theory (see also
similar situations considered in \ref\gvw{S. Gukov, C. Vafa
and E. Witten, ``CFT's from Calabi-Yau Four-folds,''
hep-th/9906070.}\ref\guk{S. Gukov,``Solitons, Superpotentials and
Calibrations,'' hep-th/9911011.}).
In the context of type IIA theory with
RR fluxes corresponding to $F$ and $G_{int}$ and $G_6$
discussed above, the superpotential
is given by
\eqn\supo{\lambda_s W=\int F\wedge k\wedge k +i\int G \wedge k+\int G_{6}}
where $k$ is the complexified Kahler class.
To see this one considers the BPS charge in the presence
of BPS domain walls which may be partially wrapped over the
CY.  For example, considering a D6 brane wrapped over
4-cycles of CY gives a domain wall with BPS tension ${1\over
\lambda_s}\int k \wedge k$
integrated
over the internal part of the 6-brane.  This in turn shifts the dual $F$
by one unit. This BPS formula should be captured by a $\Delta W$ and
we can see from the above form of \supo\ that 
the first term above precisely captures this term.
More precisely what we mean by the formula \supo\ is
the {\it worldsheet quantum corrected} formula for the kahler
forms (as is well known in the context of mirror symmetry
the mass of the
D-branes receives corrections by
the worldsheet instantons).  In particular if $t$ denotes
the complexified area of the basic $2$ cycle, then the
volumes of the $0,2,4$ and $6$ cycles are given by
$$1,t,{\partial F_0\over \partial t},2F_0-t{\partial F_0\over \partial
t}$$
where $F_0$ is the genus zero topological string amplitude.
So in particular suppose we have $N$
units of the $F$ flux through the basic $2$-cycle, where $t$ denotes
the complexified area of this 2-cycle.  Then the first term in
\supo\ is equivalent to
$$\int F\wedge k \wedge k =N {\partial F_0\over \partial t}$$
Similarly if we considered $D4$ branes wrapped over 2 cycles
and $D2$ branes with no wrappings, we deduce the existence of the
second and third term in \supo .  In particular if we denote
the fluxes of $F$, $G_{int}$, $G_6$ by integers $N,L,P$ relative
to integral 2, 4 and 6 cycles, we have
\eqn\ansupo{\lambda_s W=N{\partial F_0 \over \partial t}+i t L+ P}
Note that equation \supo\ can also be written in the form
\eqn\anfor{\lambda_s W=\int (F+i*G)\wedge k +\int G_6}
where again here by $*$ we mean the worldsheet quantum corrected
$*$ operation.

Now we come to the discussion of why turning on RR fluxes
should not modify the topological amplitudes.   We will concentrate
on genus 0 amplitudes (similar arguments can be advanced
for the higher genus amplitudes as well).  
The vector superfield
with bottom component $t$ has an auxiliary field in the superspace
of the form
$$t+\theta^2 (F+*iG)+...$$
where $F$ and $G$ are the usual RR fluxes of the internal
Calabi-Yau\foot{I have greatly benefited from discussions
with Nathan Berkovits in connection with the auxiliary
field structure of the superfields.}.
In the usual supersymmetric background they are set to zero.
Now suppose we wish to turn them on.  Suppose for example
we wish to turn on $N$ units of $F$.  Consider the topological
string amplitude $F_0(t)$.  We claim that this already 
yields the correct structure for the
generation of $N=1$ superpotential
precisely if $F_0$ {\it is unmodified in the presence of
RR flux}.  To see this note that using the expansion
of $t$ in terms of the RR field strength auxiliary fields
we have
$$\int d^4\theta F_0(t)=\int d^2 \theta N{\partial F_0 \over 
\partial t}$$
which is exactly the expected answer if $F_0$ is unmodified.
Similarly turning on the $G_{int}$ flux and using
\anfor\ we see that the term in \supo\ involving
$G_{int}$ will also have the correct structure 
if $F_0$ is unmodified.  

There is another auxiliary field in the vector
multiplet which come from the
NS-NS sector which is relevant for us.  This corresponds to the field
strength associated with the lack of integrability of complex
structure.  In particular if we write ${\overline D}={\overline
\partial}+A\partial$, where $A$ is an anti-holomorphic one form
taking values in the tangent bundle,
then 
$$ {\overline D}^2=({\overline \partial}A +[A,A])\partial={\cal 
F}\partial$$
where ${\cal F}$ is an anti-holomorphic 2-form with values in the
tangent
bundle which is equivalent, by lowering the vector index by the
three form, to a $(2,2)$ form.
If this is non-vanishing it also corresponds
to making the $(3,0)$ form in the CY not to be annihilated by
${\overline \partial}$.    These
turn out effectively to add to $F$ and $G_{int}$ complex
pieces of the form $iF^{NS}/\lambda_s$ and $iG^{NS}_{int}/
\lambda_s$. A similar $NS$ auxiliary field gives rise effectively to the
complex part of $*G_6$.\foot{Turning these fields on
is mirror to turning on $H_{NS}$ on the mirror CY.}  
In other words, even with these
fields turned on, the formula \ansupo\ remains correct
but now $N,L,P$ also include imaginary pieces of the form
$iN_I/\lambda_s,iL/\lambda_s,iP/\lambda_s$. We will continue
to denote the superpotential as \ansupo\ and keep in mind
that $N,L$ and $P$ can have complex pieces given by an integer
over $\lambda_s$.

Turning these vevs on breaks
the $N=2$ supersymmetry to $N=1$.  The field $t$ is
now the bottom component of an $N=1$ chiral multiplet
whose  auxiliary field descends from
 another auxiliary field (which also comes from the NS sector)
in the original $N=2$ multiplet which is not turned on.

Note also that the higher genus topological amplitudes also
give rise to certain $N=1$ superpotential terms
when the auxiliary field of the $N=2$ multiplet $t$
takes a vev.  In particular with $N$ units of RR flux for $F$ we get

\eqn\anan{\int d^4\theta \ {\cal W}^{2g}F_g(t)\rightarrow N\int d^2\theta
\ {\cal W}^{2g}
{\partial F_g \over \partial t}}
where we continue
to denote by ${\cal W}_{\alpha \beta}$ the reduction of the $N=2$
multiplet to an $N=1$ multiplet with the self-dual part of the graviphoton
field strength as its bottom component.

  So in conclusion we have learned that
the topological string amplitudes are not sensitive to turning
on RR field strengths, but they are useful in determining the 
superpotential terms that will be generated once certain RR and NS fields take 
a vev. This is captured by equation \ansupo .

\subsec{Open Topological Strings and N=1 amplitudes in 4d}

In the A-model, the open topological string corresponds
to studying holomorphic maps from worldsheet with boundaries
to the target space where the boundary lies on a 3-dimensional
Lagrangian subspace of the CY, i.e. the 3-dimensional topological
version of D-brane \wittcstop .
Moreover, it was shown that topological string field theory in this
case is just the Chern-Simons gauge theory on the corresponding
Lagrangian submanifold (possibly corrected with non-trivial
worldsheet instantons). The implications of these theories
for superstring amplitudes has been studied as well.  In particular
if we consider type IIA superstring in the presence of a CY with
$N$ D6 branes wrapping a lagrangian 3-cycle of CY and filling
the rest of the spacetime we get an $N=1$ gauge theory with $SU(N)$ gauge
group in 4d.  Then it was shown in \bcovii\ that
for example the genus 0 open
topological string amplitudes 
 compute corrections of the form
\eqn\impc{\lambda_sW=
\sum_h \int d^2\theta\ F_{0,h}[Nh \ S^{h-1}]+\alpha S+\beta}
where $F_{0,h}$ is the partition function of the topological 
string at genus $0$ with $h$ holes, and
 $S=\lambda_s Tr W^2$ where $W_{\alpha}$ 
is the chiral superfield with gaugino
as its bottom component\foot{The coefficient of $Nh$ in front
arises because, as discussed in \bcovii , we have to choose $h-1$
holes to put the $tr W^2$ fields and this can be done in $h$ ways
and also the trace over the hole without a field gives a factor of $N$.
Note also that $Tr W^2$ which is a fermion bilinear
is nilpotent, in the sense that $(Tr W^2)^k=0$ for $k>N^2$.
It is relevant for us precisely because we are considering
a large $N$ limit.  Thus in the large $N$ limit
the gaugino bilinear can even have a classical vev.}. 
Here we have shown explicitly the contribution coming
from  $h=2$ in the form
of
$\alpha S$.  This term, coming from annulus, is typically divergent,
signifying the RG flow of the coupling constant of the gauge theory
and needs regularization.  Also we have added a constant $\beta$
to remind us that we cannot fix that from open topological string 
considerations.
  Here $h$ is the number of holes on the sphere
and $F_{0,h}$ denotes the topological amplitude
on the sphere with $h$ holes.  If the target space has
some Kahler moduli $t$ they will correspond to chiral fields
in the $N=1$ theory in 4 dimensions and
 $F_{0,h}$ will depend on $t$.  The case of $h=1$ in the above
formula was recently discussed in \ref\douget{I. Brunner,
M.R. Douglas, A. Lawrence and C. Romelsberger, ``D-branes on
the Quintic,'' hep-th/9906200.}\ref\kach{S. Kachru, S. Katz,
A. Lawrence and J. McGreevy, ``Open String Instantons and Superpotentials,''
Phys. Rev. {\bf D62} (2000) 0260001.}\oov .  Some superstring implications of higher
genus open topological strings, i.e. 
$F_{g,h}$ with arbitrary $g$, has also been noted in \oov .   In particular
they compute terms of the form $\int d^2\theta
\ F_{g,h}\  {\cal W}^{2g} [Nh S^{h-1}]$.

Let us define the open topological string amplitude
summed over all holes at a fixed genus by
$$F^{open}_g(r)=\sum_h F_{g,h} r^h.$$
Then, for example the genus 0 open topological string amplitude
computes the following
correction to the superpotential
$$\lambda_s W= N\int d^2\theta \ {\partial F^{open}_0(S)\over \partial S}
+\alpha S+\beta $$
This is strikingly similar to the form obtained in 
\ansupo\ in the context of closed topological string amplitudes.
Similarly the higher genus correction computes terms of the form
$$N\int d^2\theta\  {\cal W}^{2g}{\partial F^{open}_g(S)\over \partial
S}$$
which is also similar to the higher genus correction obtained in the
closed string context in the presence of flux \anan .
The main difference being that $S$ is an operator for the open
string amplitudes but $t$ is a parameter in the closed string setup.
Nevertheless we will see in
 the next section  why this is not an accidental
similarity and provides the superstring interpretation
of the duality found in \gopv , when $S$ takes an expectation value
equal to $t$.

\newsec{Embedding Large $N$ Topological String duality in Superstrings}
Consider Type IIA strings in a non-compact CY 3-fold geometry
of the form of the conifold times the Minkowski space $M^4$:
The internal geometry is given by
$$f=x_1^2+...+x_4^2=\rho$$
where each $x_i$ parameterizes ${\bf C}$.  The real subspace of the
above geometry is $S^3$ (for real $\rho$) and the imaginary
directions sweep the cotangent direction of $S^3$.   The volume
of $S^3$ in string units is given by $\rho$ (here we are taking
the canonical 3-form $\Omega=\prod dx_i/df$, which scales
as $\rho$ to give the volume).
Thus
symplectically the conifold is $T^* S^3$.  Consider $N$
 D6 branes wrapped over the $S^3$ of the conifold
and filling the rest of the spacetime.  On the uncompactified
worldvolume of the D-brane we have an $N=1$ supersymmetric
$SU(N)$ gauge theory.  Note that to leading order
the action on the uncompactified worldvolume
of the D-branes is given by the superpotential
\eqn\fitr{{1\over \lambda_s}\int d^2\theta \ S Y}
where $S=\lambda_s Tr W^2$ and  where
$Y$ denotes the $N=1$ chiral superfield with its bottom
component given by $iC+{\rho \over \lambda_s}$, where $C$
is the vev of the 3-form gauge field on IIA (normalized
with periodicity $2\pi$) and plays the role of the theta
angle for the gauge theory and $\rho $ denotes the volume of the $S^3$.

The choice of this type IIA geometry is based
on the desire to utilize the topological open/closed string
duality.  In particular
 as discussed in
the previous section the open topological
string in this case computes corrections to the superpotential
of the form
$${N\over \lambda_s}\int d^2\theta \ {\partial_S}F^{open}_0(S)$$
The topological A-model is insensitive to complex structure.
In particular $F^{open}$ is independent of $\rho$ except
for a linear terms in $S$ (coming from the annulus) written in
\fitr , which is related to the ambiguity
of open topological string at the level of annulus. There
is also a divergence of the annulus amplitude corresponding
to the running of the gauge coupling constant, which, 
in the regularized form, can be
viewed as addition of a linear term in $S$. 
The corrections above to the simple $YS=YTr W^2$ involve
higher dimension operators (more powers of $S$) and are
captured by the open string amplitude which
coincide with the large $N$ expansion of the
Chern-Simon amplitudes on $S^3$ .  Note also that the fact
that they are independent of $Y$ implies that they survive
no matter what the size of the $S^3$ is. 

Now we wish to consider the limit where we consider the
$N\rightarrow \infty$ limit keeping $N\lambda_s$ fixed.  In this
limit, the analog of `t Hooft coupling for the gauge system
is given by 
$${1\over Ng_{YM}^2}\rightarrow {Y\over N\lambda_s}.$$
which remains fixed in this limit. 
We would like to consider the gravity dual of this gauge system.
In the spirit of AdS/CFT correspondence we will have to
consider the near horizon geometry.  What the precise notion
of ``near'' horizon geometry in this case should be is more subtle
because the expectation value of $Y$ undergoes an RG flow, as noted above,
and it will depend at which scale we are probing it. In other words
we have to readjust the size of $Y$ depending on how close we are
approaching the branes.
 The limit
should be such that the $S^3$ has zero size when we probe it in the UV
of the gravitational side but finite size in the IR. 
To avoid such subtleties we try to look for a consistent
gravitational background which the branes create.  In particular
we should find an $S^2$ of finite size emerging,
 surrounding the $S^3$, with the D-branes completely
disappeared and replaced by the corresponding fluxes.  In the case
at hand, since we have N $D6$ branes wrapping the $S^3$ in the
geometry after transition
we should get $N$ units of the
 2-form RR flux $F$ through the
dual $S^2$.  We will now turn to studying this geometry.

\subsec{Type IIA Superstring on the blown up geometry}
We thus seek the dual large $N$ stringy
 description of the above gauge
system, in the form of the Type IIA background
with the blown up conifold geometry, i.e. the geometry
corresponding to $O(-1)+O(-1)$ bundle over ${\bf P}^1$, with
$N$ units of 2-form $F$ flux through ${\bf P}^1$.  However we
must also have internal $4$-form and $6$-form fluxes (in the
form of NS and RR fields discussed before).

That there should be an NS $4$-form flux 
 corresponds to the fact that the size of the
$S^3$ is changing (i.e. that $\Omega$ is no
longer closed and $\rho= \int_{S^3} \Omega$ changes), inducing a running
of the gauge coupling constant.    Moreover to preserve
$N=1$ supersymmetry for a finite value $t$ of the
complexified area of the blown up ${\bf P}^1$
we need both 4-form as well as 6-form fluxes.
In fact, as discussed in the previous section and summarized
in equation \ansupo\ we have a superpotential
of the form
\eqn\supy{W=N{\partial_t}F_0(t)+it L +P}
where for the geometry at hand $F_0(t)$ is, up to a cubic
polynomial, the tri-logarithm
function, given by
\eqn\prepo{F_0(t)={1\over 6}t^3-\sum_{n>0}{e^{-nt}\over n^3}+P_2(t).}
(where $P_2(t)$ is a polynomial of order $2$ in $t$ and is somewhat
ambiguous).  Similarly for higher genus $F_g$ we have
$$F_g(t)={B_{2g}\over 2g(2g-2)!}\sum_{n>0} n^{2g-3}e^{-nt}+{B_{2g}B_{2g-2}
\over 2g(2g-2)(2g-2)!}\qquad g>1$$
$$F_1(t)={t\over 24}+{1\over 12}{\rm log}(1-e^{-t})$$
where $B_{2g}$ denotes the Bernoulli numbers.  The terms
involving $e^{-nt}$ in the above formula reflects the corrections
due to worldsheet instantons wrapping $n$ times over the ${\bf P}^1$.

The content of the duality obtained in \gopv\ is that
$$F_g^{Open}(S)=F_g(t)$$
for all $g$ if we set $S=t$.  We now try to interpret this statement
in the superstring context.
For this we need to study solutions to the gravitational equations.

Typically in physics and mathematics when we try to solve
some system of equations, there are topological obstructions
that have to be shown to be absent.  Once they are shown to be absent
then a solution exists.  For example, when we are trying to find
Ricci-flat metrics on Kahler manifolds we need the first chern
class of the manifold to be zero.  In fact this is also sufficient
for being able to find a Calabi-Yau metric.    Of course explicit
solution for the metric has not been possible in almost all cases
and in fact the Ricci-flat metric is only an approximate
metric which gives
rise to a conformal worldsheet theory.  In a sense the
topological condition, guaranteeing the existence
of the solution is more fundamental than the solution itself.

Now we come to the case at hand.
To preserve $N=1$ supersymmetry we need $W=dW=0$. Once these
are satisfied, we expect physically that there must be a
solution to the gravitational
system.  In fact a very similar example with the same number
of supercharges (namely 4) was already studied from this point
of view.  Namely if we consider M-theory on Calabi-Yau 4-fold
with G-flux turned on, the gravitational equations have been studied
 in \ref\beker{K. Becker and M. Becker,
 ``M-theory on Eight-Manifolds,'' Nucl. Phys. {\bf B477}
 (1996) 155.}.  The topological conditions they find for the existence
of the gravitational solution has been shown to be identical
to the condition that $W=dW=0$ \gvw .

 Of course the low energy gravitational equation in the present
 case can also be studied similar to what 
 was done in \beker\ and will involve
warped geometries mixing the spacetime with the Calabi-Yau.
 Even though solving
the gravity equations would be interesting,
we have to remember that due to worldsheet instantons
wrapping the ${\bf P}^1$ there are important corrections
to the gravity equations, and so at best we can trust
the low energy gravity description in the limit of large $t$.
Nevertheless, as already noted above, the superpotential terms
{\it including } the corrected string geometry,
can be incorporated to all orders in the $W$ which is computable
by topological string amplitudes.

Before even solving the conditions $W=dW=0$ we can already
interpret the duality of \gopv\ in the superstring context.
If we compare the equation \supy\ with the superpotential
given in the gauge theory side namely $W=N{\partial_S}F_0^{open}(S)+\alpha S+
\beta$ we see that they are identical in form with
an appropriate identification of $\alpha$ and $\beta$
with $L$ and $P$.  Therefore, since the vacuum in the gauge theory
side, as well
as the moduli in the gravity side 
correspond to $W=dW=0$, and $W$ has the
same form for the gauge as well as the gravitational
system, this will identify 
\eqn\fundeq{\langle S\rangle =\langle \lambda_s Tr W^2 \rangle=t\qquad
F^{open}_g(S)=F_g(t)}
Thus the condition of vacuum configuration which sets $\langle
S\rangle =t$ also translates the duality found in \gopv\ to the
match between amplitudes in the gravity side and the gauge theory side 
to all orders in $1/N$ at least as far
as superpotential terms are concerned!

Note that the idea that $\langle S\rangle
\not =0$, i.e. that we have gaugino
condensation, is very natural for the open string
theory
under discussion as it does have an $N=1$ Yang-Mills theory associated
with it.  Part of the above check involves,
on the gauge theory side, the statement that gaugino
condensation 
generates superpotential terms captured through topological open string
amplitudes and in fact this was already pointed out in \bcovii .  Of
course
here we have a more refined gauge theory than just the $N=1$
Yang-Mills theory and in particular we have in the open
string system, also the higher
dimension operators present, which are 
captured by the topological string amplitude.  At any rate, the result
of \gopv\ strongly suggests not only the existence of a large $N$
duality involving this $N=1$ brane system with this
closed string background, but also that the gaugino condensation
takes place.

We now come to finding
 solutions to the equations $W=dW=0$ on the gravity
side, which should guarantee the existence of a solution.
There are four parameters under control: the modulus $t$ and
the fluxes $N,L,P$.  The two equations $W=dW=0$ 
$${\partial_t W}=0 \rightarrow N F_0''+iL=0 \rightarrow L=iN F_0''$$
$$W=0\rightarrow P=-NF_0'+ NtF_0'' $$
imply that
two of these four quantities are fixed in terms of the other
two. 
The $N$ is of course fixed for us by the number of 
D6 branes.  As is clear from our description of the dual
gauge system the choice of a shift in $L$ is related to a shift
in the bare coupling constant of the gauge system.  In particular
in order to agree with the bare coupling constant of the gauge theory
$iL=\rho/\lambda_s $, where $\rho$ is the volume of the $S^3$ where the D6 branes
are wrapped around.
 Thus the value of $t$ (and also of $P$) is fixed   and
from \prepo\ and
$\partial_t W=0$ the solution for $t$ is given by
\eqn\ident{[c(e^t-1)]^N= {\rm exp}(-\rho /{\lambda_s})}
The constant $c$ depends on the ambiguities hidden in the $P_2(t)$.
As we will argue from the dual gauge theory description,
it should be fixed in our case (by a suitable regularization
of the one loop divergence of the gauge theory) to be $c\sim N \lambda_s$.

 Next we turn to the
question of how the dynamics of the gauge system is reflected
in the $W$ and the other superpotential terms.
What do we expect for the dynamics of the $N=1$
supersymmetric theory living on the D6 brane? If we ignore
the higher powers of $S$ in the superspace integral, i.e.
if we ignore the higher order operators,
 as already
discussed, the leading term with the lowest number of derivatives,
is given by the superpotential term
$${1\over \lambda_s}\int d^2\theta\  SY$$
where $S=\lambda_s Tr W^2 $ and $Y=iC+{\rho\over \lambda_s}$ where
$\rho$ denotes the size of the $S^3$ in the string frame.  In the
usual geometric engineering of standard $N=1$ gauge theories,
and in particular the ones discussed in 
\ref\oogvaen{H. Ooguri and C. Vafa,
``Geometry of N=1 Dualities in Four Dimensions,'' Nucl.
Phys. {\bf B500} (1997) 62.}\
 one considers
the limit where $\rho$ is large, in which case the field $Y$ gets
demoted to a parameter in the lagrangian (the corresponding D-term
involving $Y{\overline Y}$ becomes very large).  However here
we are not necessarily 
interested only in a regime where $Y$ is very large.
In other words we consider
the field $Y$ to be a  dynamical field.  Thus we have a non-standard
$N=1$ supersymmetric gauge theory
 with its coupling constant as a dynamical field. 
Even though it is somewhat unconventional, as we will now argue
some of the basic features of this theory are similar to that of $N=1$
QCD, in the limit where we ignore the higher derivative
terms of the form $\int d^2\theta [tr W^2]^k$.
In other words, if we consider the field space where $S=tr W^2 <<1$
(in string units) we have a theory which is more or less similar
to $N=1$ supersymmetric Yang-Mills theory.  Even though we do
not have to restrict our attention only to this limit, and the duality
with gravity holds regardless of which field configuration in $S$ we
consider, it is first instructive to consider the small $S$ region
to gain intuition for what this theory is.

In the dynamics of $N=1$ supersymmetric gauge theory,
a prominent role is played by instantons.  Here a similar
effect exists:
 In particular if we consider  Euclidean D2 brane instantons
wrapping the $S^3$ the superpotential gets corrected.  Moreover this
can also be viewed as point-like instantons for the $SU(N)$ gauge theory.
  To have the
right number of fermionic zero modes to lead to a chiral
superspace potential we need $1/N$-th of this instanton.  Since
the action for this instanton is $e^{-Y}$, the term that can appear
in the action is $e^{-Y/N}$.  The coefficient in front of it should
be of order $N^2$ (as argued in \ref\witln{E. Witten, 
``Branes and the Dynamics of QCD,'' Nucl. Phys. {\bf B507}
(1997) 658.}).
So we must have the effective superpotential given by
\eqn\sefo{W=\int d^2\theta ({1\over \lambda_s}SY+iN^2 \alpha e^{-Y/N})}
where the constant $\alpha $, by a shift in $Y\rightarrow Y+Y_0$,
can be identified with a shift in the
bare coupling constant of $S$, i.e.,
\eqn\mena{\alpha =e^{-Y_0/N}}
  (The choice of $\alpha$
is also related to how we regularize the one-loop divergence which
corrects the action with a term $a\int d^2\theta\  S$). 

This effective superpotential has the same structure as that
encountered in the proof of mirror symmetry in 2 dimensions
\ref\hv{K. Hori and C. Vafa, ``Mirror Symmetry,'' hep-th/0002222.}.
  This same superpotential structure was encountered
in \oov\ in the context of $N=1$ domain walls in 4d, which we
will also need in this paper, and we will discuss further below.
Notice that here since $Y$ is a dynamical field, we can integrate it out
 by setting
$$\partial_YW=0\rightarrow{1\over \lambda_s} S=iN\alpha e^{-Y/N}$$
which leads to 
$$Y={\rm log}({S\over iN \alpha \lambda_s})^{-N}$$
plugging it back to the superpotential gives the effective
superpotential for $S$:
$$W_{eff}(S)={1\over \lambda_s}[S{\rm log}({S\over iN\alpha 
\lambda_S})^{-N}
+NS]$$
This is the familiar effective superpotential expected for the
gaugino bilinear $S$ in the standard $N=1$ supersymmetric gauge theory.
Indeed setting $\partial_SW=0$ leads to
\eqn\namit{[{S\over iN\alpha \lambda_s}]^N=1\rightarrow S= i N\alpha \lambda_s
e^{2\pi
il/N}=iN \lambda_s e^{(-Y_0+2\pi i l)/N}}
Note that we see the $N$ vacua of $SU(N)$ Yang-Mills, in the
standard way.  

 Let us compare the vev we found for $S=\lambda_s Tr W^2$ with the gaugino
condensate
for standard $N=1$ Yang-Mills, which is of the form
$$Tr W^2=iN \Lambda^3 e^{[{-1\over N g_{YM}^2}]+{2\pi i l\over N}}$$
In comparison with what we have above, note that this is in perfect
agreement with \namit , where $\Lambda $ corresponds to the string scale
and $1/g_{YM}^2\rightarrow Y_0$.

Note that the effective superpotential we have found for $S$,
for small $S$, also follows from either the open topological
string amplitudes in the limit $S\rightarrow 0$,
or the dual closed topological string amplitude in the limit
$S=t\rightarrow 0$ which is given by
$$F_0(t)\rightarrow -{1\over 2}t^2{\rm log}t+at^2+bt+c$$
and so
$$W(S)={1\over \lambda_s}[N\partial_S F_0(S)+\alpha S+\beta]=
{1\over \lambda_s}( S{\rm log} S^{-N}+N\cdot {\rm const.} S+
N\cdot {\rm  const.})$$
in perfect agreement with expectations based on the gauge theory
analysis as well as with the contribution of the
Euclidean D2 brane instantons
in the string context.  This comparison with gauge theory 
and recalling that $t\leftrightarrow \lambda_s Tr W^2$,
also
fixes the value $c$ in \ident\ to be $c\sim N\lambda_s$.
Note that the choice of $\alpha$, the
linear terms in $S$,  on the gravity side is controlled by
the 4-form fluxes dual to the ${\bf P}^1$, as discussed before.

Having discussed the geometry of the vacua of $N=1$ theory,
we now turn to another important feature of $N=1$ theories, namely
the domain walls interpolating between various vacua.

\subsec{Domain Walls}
$N=1$ Yang-Mills theory admits BPS domain walls
interpolating between various vacua.  As noted in 
\witln\ at large $N$ they
behave as D-branes for QCD string.  In particular
their tension is of the order of $N$.
 Since in the
present context the QCD string is realized by the fundamental
string, ordinary D-branes of string theory should play the
role of domain walls.  This is indeed the case:  On the
gravity side we have a blown up ${\bf P}^1$.  If we
consider D4 branes wrapped over ${\bf P}^1$ they correspond
to domain walls.  Their tension goes as 
$$T\sim {1\over \lambda_s}|t|=N {|t|\over {N\lambda_s}}$$
As discussed before $|N\lambda_s | \sim |t|$ so we obtain
the expected behavior.  For the QCD domain wall the
phase of the $S$ field should change as we go from one vacuum
to another.  In particular it should shift by ${\rm exp}(2\pi i/N)$
for domain walls interpolating adjacent vacua.  Let us see how
this is realized in the gravity setup.  Since
we have identified the domain wall with $D4$ brane wrapped over
$S^2$ we should note that the value of the $G$ flux shifts as we cross
the domain wall.  Consider in particular the imaginary part of the
$Y$ field introduced earlier, which was identified with
$$ImY=C_{S^3}$$
i.e. the vev of the $C$ field along the $S^3$.  We now discuss how
this changes from the left-side of the domain wall to the
right-side.  Since the G flux should be equal to one for the D4 brane, 
it
implies that $Im Y$ should shift by $2\pi $, i.e.
$$Y\rightarrow Y+{2\pi i}$$
as we go across the domain wall. 
In fact we can find the geometry of the BPS domain walls
by the usual technique of the LG theory in 2d with $N=2$ susy
\ref\cva{S. Cecotti and C. Vafa, ``On Classification of $N=2$
Supersymmetric Theories,'' Comm. Math. Phys. {\bf 158} (1993) 569.}.
  In fact for the case at hand
similar BPS domain walls were considered in
\ref\hiv{K. Hori, A. Iqbal and C. Vafa, ``D-branes and Mirror
Symmetry,'' hep-th/0005247.}.
  These domain walls also featured in the
discussion of $N=1$ generation of superpotential in 
\oov .
Note that since we have
$$S=\lambda_s N {\rm exp}(-(Y+Y_0)/N)$$
 this implies that the phase of $S$ changes by ${\rm exp}(-2\pi i/N)$
as expected.
Of course this is suppressed at large $N$, in
agreement with the fact that classically the wrapped D-brane
does not change the value of $t$.

It is also easy to see from the form of the action \sefo\ that
the BPS tension, which is given by $\Delta W$ is given by
$$\Delta W={1\over \lambda_s}{S \Delta Y}$$
Since $S$ is identified with $t$, this corresponds to 
$$\Delta W={2\pi i\over \lambda_s}t$$
as expected for the tension of the BPS wrapped $D4 $ brane.

\subsec{Subleading Corrections in the $1/N$ Expansion}

So far we have concentrated on the interpretation of the leading
corrections in large $N$.  In the context of topological strings also
the subleading terms to all orders in $1/N$ were found to agree
between the Chern-Simons gauge theory and the closed topological
string expansion.  What is the interpretation of these higher terms
for the gauge theory system?

In the limit of small $t$ the topological string amplitudes is given
by
$$F(t)=\sum_g F_g \lambda_s^{2g-2} t^{2-2g}$$
where $F_g={B_{2g}\over 2g(2g-2)}$ and $B_{2g}$
are the Bernoulli numbers ($F_g$ turns out to be equal
to the Euler characteristic
of the moduli space of genus $g$ Riemann surfaces). In this limit
the topological string partition function coincides with that
of non-critical bosonic strings on a circle with self-dual
radius (this connection is well understood and will be reviewed
in section 5).  The $N=2$ amplitude that this computes
is given by 
\eqn\allor{\int d^4\theta \ {\cal W}^{2g} 
F_g t^{2-2g}=g R^2 F^{2g-2} F_gt^{2-2g}+...}
This correction has been physically understood
 by considering turning on
constant graviphoton field strength in the Minkowski
space and computing the effect of wrapped D2 branes on ${\bf P}^1$ 
to the $R^2$ term \ref\annaet{I. Antoniadis,
E. Gava, K.S. Narain and T.R. Taylor,
``N=2 Type II-Heterotic Duality and Higher
Derivative F-terms,'' Nucl. Phys. {\bf B455} (1995) 109.}.  In the present
context the wrapped $D2$ branes correspond to the baryon vertex,
as in the usual AdS/CFT correspondence
\ref\witbar{E. Witten, ``Baryons and Branes
in Anti de Sitter Space,'' JHEP {\bf 9807} (1998) 006.}.  The Baryon fields
 are charged under the graviphoton
field with charge proportional to their BPS mass $t$. Thus
turning on graviphoton $F$ effectively turns on a background 
field strength for $F_v$, i.e. the $U(1)\subset U(N)$ living on the D6 branes,
which can
be identified with a global $U(1)$ symmetry (the `Baryon number'
symmetry).  
Let us try to see how this can come about from the gauge theory side.

On the worldvolume of the $D6$ branes we have terms of the form
$$\int_{R^4} [G_4+F\wedge F_v](\int_{S^3}[CS(\omega )-CS(A)])$$
where $\omega$ denotes the spin connection on $S^3$ and $A$ is
the internal gauge field on $S^3$.
This term arises (by integrating by parts) from
the usual
inducement of brane charge by gravitational and gauge curvature
on the brane (see \ref\mwd{D. Diaconescu, G. Moore
and E. Witten, ``A Derivation of K-Theory from M-theory,'' hep-th/0005091.}\
and references therein).  Thus shifting $F$ effectively shifts $F_v$
\foot{Note that if we change the $G_4$ flux this is equivalent
to turning on an internal Chern-Simons action for the supersymmetric
system on the brane.  It should be possible to derive directly
the relation
between generation
of superpotentials on the brane and the Chern-Simons theory on $S^3$
from this fact.}.

There is another term that is also generated from \allor\
when we recall that $t$ has some auxiliary field turned on.  In particular
this gives rise to the term
$$N\int d^2\theta \ {\cal W}^{2g} \partial_t F_g(t)=
N\int d^2\theta \ F^{2g} \partial_t F_g(t)+...$$
Recalling that in the gauge theory setup $t$ is replaced by
$S$, the gaugino bilinear superfield, the above term corresponds
to the superpotential term
$$\lambda_s W=N\int d^2\theta \ F^{2g} \partial_S F_g(S)=-N
\int d^2\theta  {B_{2g}\over 2g} F^{2g} S^{1-2g} $$
So turning on the (self-dual) graviphoton field strength 
in 4 dimensions deforms the superpotential.  What is the gauge theory
interpretation of this?  As noted above turning on $F$ 
has the effect of turning on the field strength $F_v$
in the $U(1)\in U(N)$, 
which is also
equivalent to turning on $B$ field in spacetime.  Thus this seems
to be related to considering the non-commutative version of the above gauge
system \ref\seibwit{N. Seiberg and E. Witten,
``String Theory and Non-commutative Geometry,'' hep-th/9908142 .}.
  In particular considering a self-dual non-commutativity
  in spacetime
presumably generates a superpotential, as is predicted from the above
formula.  Note that this is consistent with the fact that in the UV
where $S$ is smaller this modification
of the superpotential is a more pronounced effect, and it disappears
in the IR where $S$ is larger. It would be interesting to derive
this result directly in the context of the non-commutative $N=1$
Yang-Mills theory.  Moreover the dependence of the
genus $g$ partition function on the non-commutativity
parameter is identical\foot{We thank R. Gopakumar
for pointing this out to us.} to that obtained in 
\ref\ssn{S. Minwalla, M. Van Raamsdonk and N. Seiberg,
``Noncommutative Perturbative Dynamics,'' hep-th/0002186.}.
Namely, in the large $N$ expansion, there is no modification
at the level of planar diagrams, i.e. at $g=0$.  Moreover
at genus $g$ the amplitudes are expected  (when we
have a self-dual non-commutativity) to
scale (in the leading order) as ${1\over \theta ^{4g}} \sim B^{4g}
$ which in our case translates to an $F^{4g}$ dependence.
This is in agreement with the fact that $|\partial_S W|^2$ 
indeed scales as $F^{4g}$.

\subsec{More General Values of $S$}
So far, in the context
of gauge theory discussion we mainly considered the limit 
where $<S>$ is small
compared to the string scale.  However the duality we are proposing
holds for arbitrary $<S>$.
If  $<S>$ is not small, on the gauge theory side we get modification
to the form we have written above, which is computed by the
Chern-Simons theory on $S^3$. What kind of gauge theory
does this correspond to? The gravity side provides a hint:
If we consider wrapped D4 brane domain walls, we have infinitely
many species of domain walls.  The reason for this is that we
can consider the bound state of $n$ D2 branes with the D4 brane
manifested through turning on $n$ units of $U(1)$ flux
through the $S^2$ part of the worldvolume
of the D4 brane.  This can also be viewed as the effect of
changing
the B-field on the ${\bf P}^1$ by $2\pi i n$.  The effect
of such domain walls is thus shifting $t=S\rightarrow S {\rm exp}
(2\pi i/N) +2\pi i n$.  In other words we have the vev's
of $S$, not only taking values around a circle about the origin,
but also circles about $2\pi i n$ for any integer $n$.
Moreover the BPS tension for such domain walls is given by
$$\Delta W={1\over \lambda_s}(S+2\pi in)$$
  The
geometry of these domain walls can be recovered
from an enlarged field content \oov :  We can introduce one
variable $Y_n$ for each $n$, capturing the corresponding
domain wall by its shift in the argument, and consider the superpotential
\eqn\infva{W=\int d^2\theta \sum_n [(S+2\pi i n)Y_n+iN^2\alpha 
e^{-Y_n/N}]}
the domain wall with 1 D4 brane wrapped
over ${\bf P}^1$ bound to $n$ D2 branes will now correspond to shifting
$Y_n\rightarrow Y_n+{2\pi i}$. Integrating the $Y_n$'s out will
give
$$W={1\over \lambda_s}
\sum_n (S+2\pi i n){\rm log}(S+2\pi in)^{-N}+a(S+2\pi i n)+b$$
which is indeed equal to
$$W=\int d^2\theta {1\over \lambda_s}N{\partial F_0(S)\over \partial S}.$$
The variables $Y_n$ were introduced to incorporate the kinks,
but their appearance on the original gauge theory side, except
for $Y_0$ seems mysterious.  It would be interesting to see if
one can find a direct interpretation of all the $Y_n$'s.  We expect
that to be related to the possibility of doing large $SU(N)$
gauge
transformations
on the $S^3$ part of the worldvolume of D6 brane.

The higher genus corrections in the case of large $S$ are also
similar to the modification at the genus 0 case.  In particular
we get an infinite sum with $S$ replaced by $S+{2\pi i n}$.
This in particular is related to the fact that we can have a new
baryon vertex for each wrapped D2 brane with D0 brane turned on
\ref\gopvaf{R. Gopakumar and C. Vafa, ``M-theory and
Topological Strings I,II,'' hep-th/9809187, hep-th/9812127.}.  

\subsec{Adding Matter}
In the context of geometric engineering of $N=1$ supersymmetric
gauge theories realized as D6 branes wrapped around $S^3$ cycles
of CY manifolds \oogvaen\
matter can be realized as extra $D6$ branes wrapped around other $S^3$'s
intersecting the gauge theory $S^3$ along a circle (where
the vev of the Wilson line around the circle on the probe brane plays
the role of mass for the matter).  How does our
duality extend to this case?  In fact in the topological string
the duality does extend to this case \oov .  In particular
in computation of Wilson loop observable
for the Chern-Simons theory one adds extra topological
branes intersecting the original $S^3$ along a knot, and it was
shown that the closed topological string amplitudes agrees
with the expected result for knot invariants for Chern-Simons
theory.  More checks have been made in \mala\
for a large number of distinct knots.  In the context
of embedding the topological string dualities in the superstring
what this means is that the dual gravitational system will not only have
a blown up $S^2$ but will also have additional $D6$ branes
(which for algebraic knots will intersect the $S^2$ along a circle).
The fact that the topological computations agree on both
sides translates to the statement that the superpotential computations
on both sides agree and is further evidence for this duality
in the superstring context.  
Note that for each knot we obtain a different ``matter'' system
for this generalized gauge theory, which in the limit
of large $Y$ give rise to the same low energy physics, but
are distinct theories in the context of generalized gauge
theories we have been considering.  The gauge theoretic
interpretation of these results is currently under investigation
\ref\work{C. Vafa, work in progress.}.

\newsec{The Mirror Type IIB Description}

As is well known, type IIA on a CY is equivalent to type IIB
on a mirror CY.  This implies that everything we have said
above in the context of type IIA has a type IIB counterpart.

For example instead of D6 branes of type IIA wrapped around
$S^3$ we consider D5 branes of type IIB wrapped around $S^2$.  Also
turning on even-form  fluxes in type IIA is mirror
to turning on 3-form $H_{RR}$ and $H_{NS}$ flux in the type
IIB side and the superpotential that gets generated
in this context is given by 
$$W={1\over \lambda_s}\int \Omega \wedge [H_{RR}+\tau H_{NS}]$$
where $\tau$ is the complex coupling constant of type IIB,
and $\Omega $ is the holomorphic 3-form of the CY.  The above
integral can be done and yields the  formula
in terms of the prepotential of the corresponding $N=2$
theory, as discussed 
in the Type IIA case.  Note however, that the type IIB
system is simpler in that by mirror symmetry
the worldsheet instantons that were relevant in the context
of type IIA theory in computing the prepotential, are absent
for the type IIB case, and classical geometry already
captures these corrections. In particular the B-topological
theory (known as the Kodaira-Spencer theory of gravity)
simply involves aspects of complex geometry of Calabi-Yau.

  So as far as writing a classical
gravitational background, the type IIB description would be more
useful because the worldsheet instanton effects are absent.
  However, as far as the conformal theory on the
string worldsheet, i.e.
the large $N$ expansion description of the gauge system,
 the type IIA and type IIB theories
are of course identical.

In the above context we would need to know the mirror of
local CY: $O(-1)+O(-1)\rightarrow {\bf P}^1$.
The mirror of this is known and it is essentially the conifold
with one subtlety \hiv :  The conifold has only one compact
3-cycle, whereas $O(-1)+O(-1)\rightarrow
{\bf P}^1$ has two compact even
cycles, namely
$0$ and $2$ cycle.  As was noted in \hiv\ in the limit
where the Kahler class of ${\bf P}^1$ approaches zero, i.e.,
$t\rightarrow 0$, the mirror becomes effectively the conifold
(the actual mirror differs from the conifold by having some
variables being ${\bf C}^*$ variables rather than ${\bf C}$
variables)\foot{The actual mirror is given by
$x_1+x_2+x_1x_2 e^{-t}+1-uv=0$ where $x_1,x_2$ are ${\bf C}^*$
variables and $u,v$ are ${\bf C}$ variables.}.
Similar observations were made in \ref\maga{
M. Aganagic, A. Karch, D. Lust and A. Miemiec,
``Mirror Symmetries for Brane Configurations and Branes at Singularities,''
Nucl. Phys. {\bf B569} (2000) 277-302.}.  Even though in principle we can
consider the full mirror geometry, since the complex
geometry of the conifold is more familiar and better studied
we restrict our attention to this case\foot{In this limit
the internal topological theory corresponds to a
$G/G$ model on $S^2$ (coming from the holomorpohic
Chern-Simons theory on $S^2$ \wittcstop \ref\vafex{C.
Vafa, ``Extending Mirror Conjecture to Calabi-Yau with Bundles,''
hep-th/9804131.}) 
which should also be equivalent, by mirror
symmetry, to the large $N$ fixed $k$ limit of the Chern-Simons theory.
This topological theory should also be equivalent to the Penner Matrix
model.  It would be interesting to verify these equivalences 
among these topological gauge theories more directly.}.
  This will correspond
to a particular limit of our Type IIA theory, where we
consider only the small $<S>$ region.  Recall that this
was the regime where the theory retained only the
leading dimension operators in the action and led
to a theory which was similar to the standard $N=1$
supersymmetric gauge theory.

We will be brief for this case, as most of the discussion
can be literally borrowed from our discussion in the previous
section.  We start with $N$ D5 branes wrapped over the
${\bf P}^1$  in the
$O(-1)+O(-1)\rightarrow
{\bf P}^1$ geometry.  The large $N$ limit of this, in the limit
of shrinking ${\bf P}^1$ corresponds to blowing up an $S^3$
with $N$ units of $H_{RR}$ flux through the $S^3$.
Let us write the conifold geometry as
$$z_1z_2-z_3z_4=\mu$$
Then the genus 0 prepotential is given as
$$F_0(\mu)={-1\over 2}\mu^2{\rm log}\mu +P_2(\mu)$$
where $P_2(\mu)$ is an undetermined polynomial of degree
$2$ in $\mu$. Now we consider turning on fluxes:
 The mirror of
turning on $NS$ 4-form flux corresponds here to turning on $H_{NS}$
and in the cycle dual to $S^3$.  Thus as far as the
superpotential is concerned we have
$$W={1\over \lambda_s}[N\partial_\mu F_0(\mu)+M\mu ]={1\over 
\lambda_s}[-N\mu{\rm log}\mu +a
\mu + b]$$
where $M=M_1+\tau M_2$, and the discussion reduces to the
small $S$ limit of the discussion in the previous section.
 
While this paper was being prepared two papers \klst \maldn\ 
appeared which
are related to this type IIB construction.  In particular 
(among other things) they
consider the gravitational background corresponding to D5 branes wrapped
on 2-cycles of CY and their results are consistent with the superpotential
analysis here.

\newsec{c=1 Non-critical Bosonic String and N=1 Superpotentials at Large $N$}

As discussed above the type IIA or type IIB near a conifold background
with some fluxes turned on can be interpreted as large $N$ limit
of certain $N=1$ supersymmetric gauge theories.  In particular
the string expansion is equivalent to the large $N$ expansion
of a gauge theory.  Moreover certain superpotential corrections of the
gauge theory can be viewed as computations of the corresponding topological
strings in the CY background.  These are readily computed and thus
carry a large amount of information to all orders in $1/N$,
for the gauge theories in question.  In particular here
we will explain how the Type IIA superstring near the small
blow up of conifold, or equivalently the Type IIB superstrings
in the conifold geometry relate the non-critical bosonic string amplitudes
with that of the superpotential computations at the large
$N$ limit of the corresponding $N=1$ gauge systems.

It was shown in \ref\ghosv{D.
Ghoshal and C. Vafa, ``c=1 String
as the Topological Theory of the Conifold,'' Nucl.
Phys. {\bf B453} (1995) 121.}\ that the conformal theory
near the conifold is given by the same system found in 
\ref\mukv{S. Mukhi and C. Vafa,
``Two Dimensional Black Hole as a Topological Coset  Model
for Two-Dimensional String Theory,'' Nucl. Phys. {\bf B407}
(1993) 667.}\
in connection with non-critical bosonic strings on a circle of self-dual
radius. This conformal theory is that of a supersymmetric
 Kazama-Suzuki coset construction 
$$SL(2)/U(1)$$
at level $k=3$, and the relation with non-critical bosonic strings is that
the topological twisting of this system is equivalent to considering bosonic 
string propagating on a circle of self-dual radius with the fermions
of the coset model playing the role of the ghosts in the bosonic string.
This relation  between bosonic string on a self-dual circle
 and the superconformal theory of a conifold is in agreement with the fact
\ref\wittgrr{E. Witten, ``Ground Ring of Two Dimensional String Theory,''
Nucl. Phys. {\bf B373} (1992) 187 \semi E. Witten
and B. Zwiebach, ``Algebraic Structures and Differential
Geometry in 2-d String Theory,'' Nucl. Phys. {\bf B377} (1992) 55.}\
that the ground ring of the bosonic string for this background
is isomorphic to the holomorphic function on the conifold (which
is generated by $z_1,z_2,z_3,z_4$ subject to the relation $z_1z_2-z_3z_4
=\mu$) where the cosmological constant of the bosonic
string is mapped to the deformation parameter of the conifold.
Moreover the observables of the $c=1$ theory are mapped to deformations
of the conifold geometry:
$$\sum_{n}\epsilon_n(z_1,z_2,z_3,z_4)+z_1z_2-z_3z_4=\mu$$
where $\epsilon_n$ is a polynomial of degree $n$ is $z_i$.  These deformation
parameters get
mapped to states of the bosonic string which are indexed by
a representation of $SU(2)_L\times SU(2)_R$ of this system,
viewing $z_i$ as entries of a $2\times 2$ matrix $M$ with the conifold
being defined as $det M=\mu$ and where the $SU(2)_L$ and $SU(2)_R$
are realized by left and right multiplication of $M$
with $SU(2)$. In particular the degree $n$ polynomial $\epsilon_n$
decomposes into representation of spin $(j_L,j_R)=(n/2,n/2)$
 with $|m_L,m_R|\leq n/2$. Let us denote the totality of
 these parameters by $\mu_i$ (except for $\mu$).
  The bosonic string amplitudes compute
topological B-twisting of the deformed conifold\foot{Aspects of this
relation has recently been verified and certain 
results of bosonic strings have been recovered directly
using the Calabi-Yau
picture and the Kodaira-Spencer theory \ref\diva{R. Dijkgraaf and C. Vafa,
unpublished.}.}. For various aspects of $c=1$ non-critical bosonic
string see \ref\mooret{P. Ginsparg and G. Moore, ``Lectures
on 2d Gravity and 2d String Theory,'' TASI summer school 1992, hep-th/9304011.}.
 As already discussed
the genus $g$ partition function will be a function
$F_g(\mu, \mu_i)$ of these parameters deforming the conifold background.
Recall that in the gauge theory context $\mu$ is identified with $S$
and we will thus denote $F_g(S,\mu_i)$.
The topological string computes, at genus $g$, the term in the effective
action given by
$${N\over \lambda_s}
\int d^2\theta [{\cal W}^2]^g \partial_S F_g(S, \mu_i)=\int d^2\theta
 {F^{2g}}\partial_S F_g(S,\mu_i)+
...$$
What is the interpretation of this for the gauge theory?  As in
the usual AdS/CFT correspondences, we would expect that the $\mu_i$
will be related to operators on the gauge theory side, deforming the
gauge theory action by terms $\mu_i {\cal O}_i$. In fact, in the
context of 3-brane probes of the conifold \ref\klwit{I. Klebanov
and E. Witten,`` Superconformal Field Theory on 
Threebranes at a Calabi-Yau Singularity,'' Nucl. Phys. 
{\bf B536} (1998) 199.}\
aspects of such deformations for the gravity side have been studied 
in \ref\seifet{D.P. Jatkar
and S. Randjbar-Daemi, ``Type IIB String Theory on $AdS_5\times T^{nn'}$,''
Phys. Lett. {\bf B460} (1999) 281.}\ and a similar analysis
should be extendable to the case at hand.  Thus the topological
strings compute the response of the system upon such deformations.
In particular at genus $0$, turning on the $\mu_i$ modifies
the superpotential for the gaugino superfield.  Also turning on $F$
will give rise to $1/N$ correction to the superpotential. 
It would be extremely interesting to understand the source of these
corrections on the gauge theory side.

It would also be interesting to find the conformal theory associated
with the RR and NS fluxes turned on in the conifold geometry.  This
is very interesting in view of the fact that before turning fluxes
on we have an exactly solvable conformal theory given by $SL(2)/U(1)$
KS model.  It would be very interesting to find the deformation
of this theory.  It is likely to involve ingredients similar
to the ones encountered in \ref\bvw{N. Berkovits, C. Vafa
and E. Witten, ``Conformal Field Theory of AdS Background with
Ramond-Ramond Flux,'' {\bf JHEP} 9903 (1999) 018.}\ref\bezw{N.
Berkovits, M. Bershadsky, T. Hauer, S. Zhukov and B. Zwiebach,
``Superstring Theory on $AdS_2\times S^2$ as a Coset Supermanifold,''
Nucl. Phys. {\bf B567} (2000) 61.}.

\newsec{Wrapped D-branes and Compact CY}
Consider type IIA superstrings compactified on a Calabi-Yau
threefold.  In the above, we considered a situation where
we take a large number of $D6$ branes wrapped over
an $S^3$ in the CY, and taking the analog of
the near horizon geometry, to decouple the gravity,
and then proposing a dual gravity description for the gauge system.

 If  we wish to repeat what we did in the previous
sections, by considering D6 branes wrapped on some $S^3$, which
is part of a compact Calabi-Yau, and filling
the rest of the spacetime, we immediately run into a problem.
We cannot wrap a D6 brane over a 3-cycle as there would be
nowhere for the flux to go for compact internal CY.  However suppose
we consider a CY manifold with some number of $S^3$'s
and we wrap D6-branes and anti-D6 branes over them, in such
a way that the net D6 brane charge is zero.
  This is of course
a non-supersymmetric situation.\foot{We can
also consider the type IIB mirror of this
where we consider D5/anti-D5 branes wrapped
around vanishing $S^2$'s of the CY.} We expect that the branes
will eventually annihilate each other leaving us with an $N=2$
background. 
 If the $S^3$'s are rigid then this annihilation process takes
some time, because there is a potential barrier for the wrapped
D6 branes to move in the CY (i.e. there is a potential for the scalar
corresponding to moving them in the normal direction in the CY).

We now wish to apply the considerations
of this paper and propose a large $N$ dual in this context.
For the considerations of the gauge theory to be applicable
we have to consider the limit where $S^3$ has shrunk to zero size.
This is the analog of $Y_0=0$ in the formula in section 3).  What we
will find is that taking the large $N$ limit induces a transition
to a topologically distinct CY, with some fluxes turned on.  Moreover
the fluxes can disappear as in \bupo\
 leaving us with an $N=2$ supersymmetric
vacuum.  Thus the effect of the brane annihilation at large $N$ has
been to shift the background. 

Consider a Calabi-Yau with $R$ vanishing $S^3$ cycles $[C_i]$ which
span a $K<R$ dimensional subspace of $H_3$.  In other words
assume 
$$\sum Q_{ji}[C_i]=0 \quad for\quad j=1,...,R-K$$
for some integral matrix $Q$.
Let us consider $N_i$ D6 branes partially wrapped around $C_i$,
where we allow some $N_i$ to be negative, in which case we mean
the number of the corresponding anti-branes. The condition
that the net $D6$ brane flux is zero implies that
$$N_i=l_{j} Q_{ji}$$
for $R-K$ integers $l_{j}$.  Now, let us consider the limit
where the $S^3$'s are vanishingly small.  In this limit,
applying the discussion of the near horizon geometry,
we are naturally led to consider the $S^2$ blown up geometry
where the blow up parameter for the $i$-th sphere
is given by
$$t_i=i\lambda_s N_i=i\lambda_s l_{j} Q_{ji}$$
Notice that not all the $t_i$ are independent.  In particular there
are
only $R-K$ independent parameters $l_j$ which determine them.  This
is exactly as it should be for the local geometry to have a blowup\foot{
In fact this
shift in the hodge numbers can be understood from the viewpoint of 
inverse process of Higgsing 
of $U(1)^{R-K}$ by $R$ 
charged fields \ref\gms{B.R. Greene, D.R. Morrison and A. Strominger,
``Black Hole Condensation and the Unification of String Vacua,''
Nucl. Phys. {\bf B451} (1995) 109.}\
 where the charged
fields can be viewed as wrapped $D2$ branes in the blown up geometry.}.
In other words the blow up geometry is a CY with $K$ less dimension
of $h^{2,1}$ but $R-K$ more dimensions of $h^{1,1}$. 
Moreover the condition on the various Kahler classes of the
${\bf P}^1$'s is exactly the same as that found above for the $t_i$.
This gives further support for the conjecture that the large $N$ limit
of the wrapped brane-anti-brane geometry, when we have no net branes is 
equivalent to a blown
up CY with the Kahler parameters for the blown up spheres given as above.
However, here we will also have RR fluxes through the $S^2$'s.  
In this case the supersymmetry is completely broken \pols\Mich\
 by the RR fluxes.  The
fluxes can disappear as recently studied in \bupo .

It would be interesting to check this generalized conjecture
in the topological string setup:  Namely this suggests that the
topological open string amplitudes in the context of compact
CY manifolds
when there is no net topological D-brane, and when the D-branes
are wrapped over spheres is easily computable by a related
closed string theory computation on the blown up CY with different
Hodge numbers.

\newsec{Generalizations}
There are many natural generalizations of this work.  In particular
it is natural to consider transitions among topologically distinct
manifolds, going through vanishing cycles, and find a large $N$
brane system/gravity duals, where the large $N$ gauge system
will lie on one side of the transition and the dual gravitational
system will lie on the other side (this was in fact the philosophy
advocated in \gopv )\foot{If we consider a Morse function $f$ on a manifold
the critical points of it encode certain topological aspects of the manifold.
Near a critical point with $p$  positive and $q$ negative eigenvalues
for $\partial_i\partial_j f$, for $f\not= f_{critical}$ the manifold
near the critical point has the geometry of a
filled $S^{p-1}\times S^{q-1}$.  For $f>f_{critical}$ the
$S^{q-1}$ is filled and for $f<f_{critical}$ the $S^{p-1}$ is filled.
This is the general kind of transition expected for large number
of branes
replaced by fluxes.  If we consider two manifolds in the same cobordism
class, and consider a Morse function on the interpolating manifold
the above picture suggests that branes can induce the transition.
So if the cobordism classes are trivial and we have suitable
branes we can interpolate between any two manifolds in this way.}.
 In other words the large $N$ brane systems can be
viewed as inducing transitions in the background geometry.
For examples there are transitions in the CY which involve
the vanishing of certain 3-cycles and blowing up 4 manifolds, such as
Del-Pezzo manifolds. In this context it is natural to conjecture the
existence of a duality involving
a large $N$ limit of wrapped D6 branes about the 3-cycles in the context
of type IIA
with the 4 manifold resolution of the singularity on the gravity
side, with certain fluxes turned 
on\foot{ In fact there is already evidence for some such
cases based on quotienting the Chern-Simons duality on $S^3$ by
finite groups on both sides \ref\anref{R. Gopakumar, A. Klemm, 
S. Sinha and C. Vafa, unpublished.}.  For example Chern-Simons
on $S^3/Z_2$ should be equivalent to ${\bf P}^1\times {\bf P}^1$ blow
up inside a Calabi-Yau.  Some aspects of this predictions
have already been checked.}.  Or in the context
of M-theory on 4-folds it is natural to look for transitions involving
shrinking $S^3$'s and growing 4-cycles, where we consider 
a large number of $M5$ branes
wrapped over the $S^3$'s and filling the 3-dimensional spacetime,
which should be dual to the geometry involving the blowup of the 4-cycle
with some $G$ flux turned on (in fact in this context the
gravity solutions are already worked out in \beker ).

It is also natural to extend our results for the case of $SU(N)$
systems to $SO$ and $Sp$ groups by including orientifolds.
In fact it has been shown in \ref\vsin{S. Sinha and C. Vafa, to appear.}\
that the large $N$ duality of Chern-Simons theory for $SU(N)$ groups
extend to the $SO$ and $Sp$ case as well. 

Finally, the idea that studying BPS/anti-BPS systems are important for a
more fundamental understanding of basic degrees of freedom for string
theory as advocated by Sen is in line with the example we have found:
We can describe one string background in terms of 
the ground state of a different
one in the presence of D-brane/anti-D-brane systems.  In a sense, this
idea, combined with the idea that various transitions among
manifolds can be induced by large $N$ limit of brane systems,
suggests that if we start with any background in string theory, and consider
complicated enough configurations of branes and anti-branes, we
can effectively be discussing arbitrary backgrounds of string theory.

{\bf Acknowledgements:}

I have greatly benefited from discussions with N. Berkovits.
I would also like to thank M. Aganagic, M.F. Atiyah,
R. Gopakumar, K. Intriligator, S. Katz,
A. Klemm, J. Maldacena, S. Minwalla, N. Seiberg, S. Sinha
and A. Strominger for valuable
discussions.

 This research is partially supported by 
NSF grant PHY-98-02709.

\listrefs

\end